\documentclass[letterpaper,twocolumn,11pt,accepted=2024-10-15]{quantumarticle}
\pdfoutput=1
\usepackage[utf8]{inputenc}
\usepackage[english]{babel}
\usepackage[T1]{fontenc}
\usepackage{amsmath}
\usepackage{hyperref}
\usepackage{physics}
\usepackage{tikz}
\usepackage{lipsum}
\usepackage{nicefrac}
\usepackage{upgreek}
\usepackage{soul}

\begin{document}

\title{Benchmarking a trapped-ion quantum computer with 30 qubits}

\author{Jwo-Sy Chen}
\author{Erik Nielsen}
\author{Matthew Ebert}
\author{Volkan Inlek}
\author{Kenneth Wright}
\author{Vandiver Chaplin}
\author{Andrii Maksymov}
\author{Eduardo P\'aez}
\author{Amrit Poudel}
\author{Peter Maunz}
\author{John Gamble}

\affiliation{IonQ, College Park, MD 20740}

\maketitle

\begin{abstract}
Quantum computers are rapidly becoming more capable, with dramatic increases in both qubit count \cite{kim2023evidence} and quality \cite{moses2023race}.
Among different hardware approaches, trapped-ion quantum processors are a leading technology for quantum computing, with established high-fidelity operations and architectures with promising scaling.
Here, we demonstrate and thoroughly benchmark the IonQ Forte system: configured as a single-chain 30-qubit trapped-ion quantum computer with all-to-all operations.
We assess the performance of our quantum computer operation at the component level via direct randomized benchmarking (DRB) across all $\binom{30}{2}=435$ gate pairs.
We then show the results of application-oriented~\cite{IonQ_AQ20_2022, qedcPeerReviewed} benchmarks and show that the system passes the suite of algorithmic qubit (AQ) benchmarks up to \#AQ~29.
Finally, we use our component-level benchmarking to build a system-level model to predict the application benchmarking data through direct simulation.  
While we find that the system-level model correlates with the experiment in predicting application circuit performance, we note quantitative discrepancies indicating significant out-of-model errors, leading to higher predicted performance than what is observed.
This highlights that as quantum computers move toward larger and higher-quality devices, characterization becomes more challenging, suggesting future work required to push performance further.
\end{abstract}

\section{Introduction}
The state-of-the-art in quantum computing is rapidly advancing, with researchers actively pursuing multiple promising technology platforms, such as superconducting circuits \cite{kjaergaard2020superconducting}, electronic spins \cite{burkard2023semiconductor}, photonics \cite{slussarenko2019photonic}, neutral atoms \cite{henriet2020quantum}, and trapped ions \cite{bruzewicz2019trapped}.
Currently, state-of-the-art systems transit the noisy intermediate-scale quantum (NISQ) \cite{Preskill2018quantumcomputingin} regime, where contemporary quantum processing units (QPUs) have dozens or hundreds, rather than a handful, of qubits.
The level of complexity of these systems presents a variety of challenges, including control system limitations, qubit cross-talk, context dependence and qubit quality uniformity.
While it is challenging to maneuver a handful of qubits to high-fidelity operation, it requires significantly more stability and control to orchestrate system-level performance that results in high-fidelity circuit execution across many qubits.
For example, the number of distinct gate pairs in an all-to-all connected architecture scales as ${N \choose 2} \in \mathcal{O}(N^2)$, where $N$ is the number of qubits in the ion chain, and so the task of calibrating all gate pairs to achieve a consistent high performance also scales as $\mathcal{O}(N^2)$.

Alongside this challenge is an associated one: how do we judge holistic performance of a many-qubit QPU?
One option to qualify user experience is through \emph{application-oriented} benchmarks \cite{qedcPeerReviewed}.
These benchmarks, similar to benchmarking suites for classical computing, aim to test performance over a variety of artificial yet realistic problems. 
Performance on a problem includes classical pre- and post-processing steps such as compiler optimization and error mitigation, allowing refinements in these steps to give higher scores.
By analyzing computer performance over these application benchmarks, we aim to assess quantum computer quality as a user would experience it.

Though application-oriented benchmarks are useful at characterizing user experience, they are not good tools for diagnosing and fixing problems in the QPU hardware.
This is because any local problems (\emph{e.g.}, a defective qubit) in application-oriented benchmarks are effectively integrated across an algorithm. 
For spotting problematic qubits, tuning and initialization, and assessing best-case capability, \emph{component-level} benchmarks, such as randomized benchmarking, are more useful. 
It is important to note that these component-level benchmarks themselves do not tell the whole story of user experience, and must be taken together with application-oriented benchmarks.

In this work, we connect component-level benchmarks with application-oriented benchmarks.
Such connection is a natural extension for holistic characterization, but is non-trivial.
Quantum operations can have complex structure in their errors that manifest differently depending on the application being run.
We simulate a suite of application benchmarks using a simple model of our QPU constructed from our component-level benchmarking.  By comparing the results to observed data, we test the extent to which component-level performance can predict the behavior of application-sized circuits.

We apply both component-level and application-oriented benchmarks to a 30-qubit trapped-ion quantum computer.
In Forte's present architecture, the 30 qubits are all-to-all connected in a single, linear ion chain; we present detailed component-level benchmarking of all 435 possible pairs.
We show that the system passes at a \#AQ 29 level, meaning that we have acceptable (greater than 1/e in circuit Hellinger fidelity \cite{qedcPeerReviewed}) outcomes on a standard corpus of representative volumetric testing circuits out to pre-optimized two-qubit gate counts of $29^2 = 841$. 
This result is possible through a combination of gate quality, compiler optimization, and error mitigation, which we explain in-depth.
We use component-level benchmarking to simulate application-oriented benchmarks via a depolarization model of the QPU.
Overall, we find reasonable correlation between theory and experiment, but we note substantive quantitative discrepancies, and that the model predicts higher performance than what is observed in experiment.

This paper is organized as follows.
In Sec.~\ref{sec:exp_setup} we describe the setup of our system, IonQ Forte, highlighting relevant architecture details.
We then provide benchmarking results in Sec.~\ref{sec:bench_and_char}, including component-level benchmarks and application-oriented benchmarking.
There, we discuss both the raw physical performance of the device, as well as the impact of compiler and error mitigation optimizations.
In Sec.~\ref{sec:sims} we describe the construction of the QPU depolarization model and the circuit simulation of the application-oriented benchmarks, and then compare this simulation to experiment.
In Sec.~\ref{sec:conclusion} we offer concluding remarks and future outlook.

\section{Experimental setup}
\label{sec:exp_setup}
 IonQ Forte uses an in-house-designed surface linear Paul trap, which consists of separate loading and quantum operation zones, and, though it is not fundamentally necessary for qubit operations, the trap is housed in a closed-cycle cryostat reaching temperatures below 10~K. 
 Neutral ytterbium atoms are generated via laser ablation, which are then ionized (\textsuperscript{171}Yb\textsuperscript{+}) via resonance-enhanced multiphoton ionization using lasers at $399$~nm and $391$~nm and trapped in the loading zone. 
  The ions are transported to the quantum operation zone to form a long chain. 
 In the work presented here, we form a 36-ion chain in the quantum zone by sequential loading, transporting and merging of individually loaded ions. 
 Each chain-loading event typically takes less than $30$ minutes.
 Once loaded, the 36-ion chain generally persists in the quantum region, ready for information processing, for tens of hours to a few days. 
 We control the trapping electric field such that the center $32$ ions are equally spaced by approximately 3~$\upmu$m, 30 of which serve as the computational qubits. 
 Similar to Ref.~\cite{benchmark11Qubits}, we encode quantum information in the ground-state hyperfine levels, $\ket{0(1)} \equiv \ket{F=0(1), m_F=0}$ of $^2\text{S}_{1/2}$, whose preparation, gate operations, and read-out are mediated by the dipole-allowed transition to the higher energy level $^2\text{P}_{1/2}$. 
 All transverse motion in the direction parallel to the trap surface in the $36$-ion chain is laser-cooled to near the ground state, employing sequential Doppler and EIT cooling~\cite{feng_2020EITcooling}. 
 The laser cooling process takes approximately $3$~ms.
 Then, all the ions are optically pumped to the $\ket{0}$ state before gate operations. 
 To read out the quantum state of qubits, a laser beam resonant with the $\ket{1} \leftrightarrow~ \ket{F=0\text{,}~m_F=0}$ of $^2\text{P}_{1/2}$ transition illuminates the entire ion chain. 
 The photons scattered from each qubit are collected by an in-vacuum high-numerical-aperture (NA) lens and directed to an individual multi-mode fiber of a fiber array. Each fiber is attached to a photomultiplier tube (PMT) for photon counting, which allows for simultaneous quantum state read-out of all the ions in the chain.
 The average state preparation and measurement (SPAM) error is measured to be $0.5\%$ on each qubit across the entire ion chain.

\begin{figure}[tb]
  \centering
  \includegraphics[width=\columnwidth]{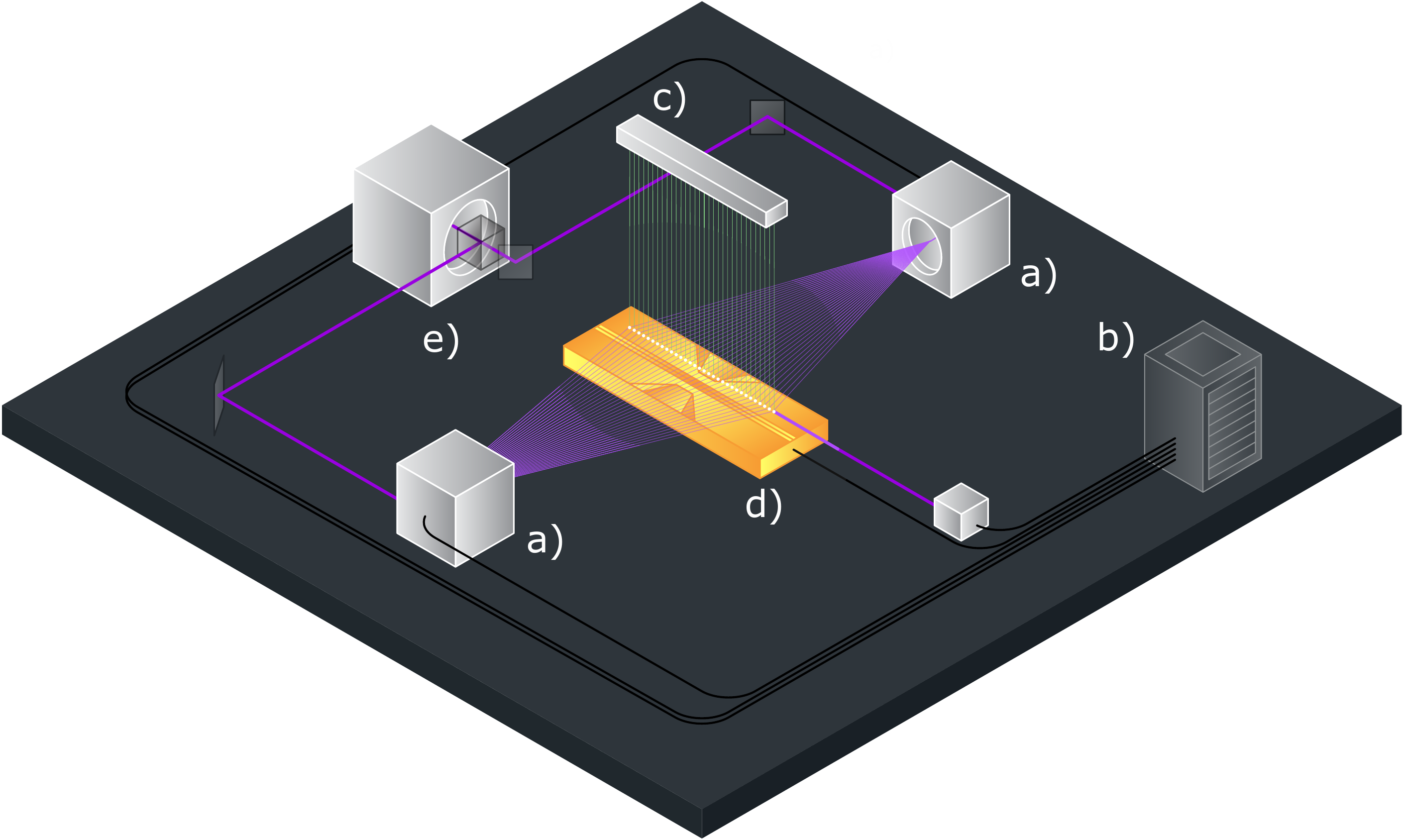}
  \caption{
  Schematic of IonQ Forte, showing the ion chain axis with respect to addressing Raman beams and imaging. 
  The two AOD devices directing light onto each ion are represented as the objects labeled (a); these devices are controlled by an embedded control system as shown by (b). 
  The ions are imaged onto individual fibers from above as depicted by (c). 
  The ions are held above the surface of an in-house designed surface ion trap (d). 
  The output of the Raman source laser (e) is split into the two arms of a counter-propagating beam pair before being modulated by each AOD. 
}
  \label{fig:forte}
\end{figure}

\begin{figure}[tb]
  \centering
  \includegraphics[width=\columnwidth]{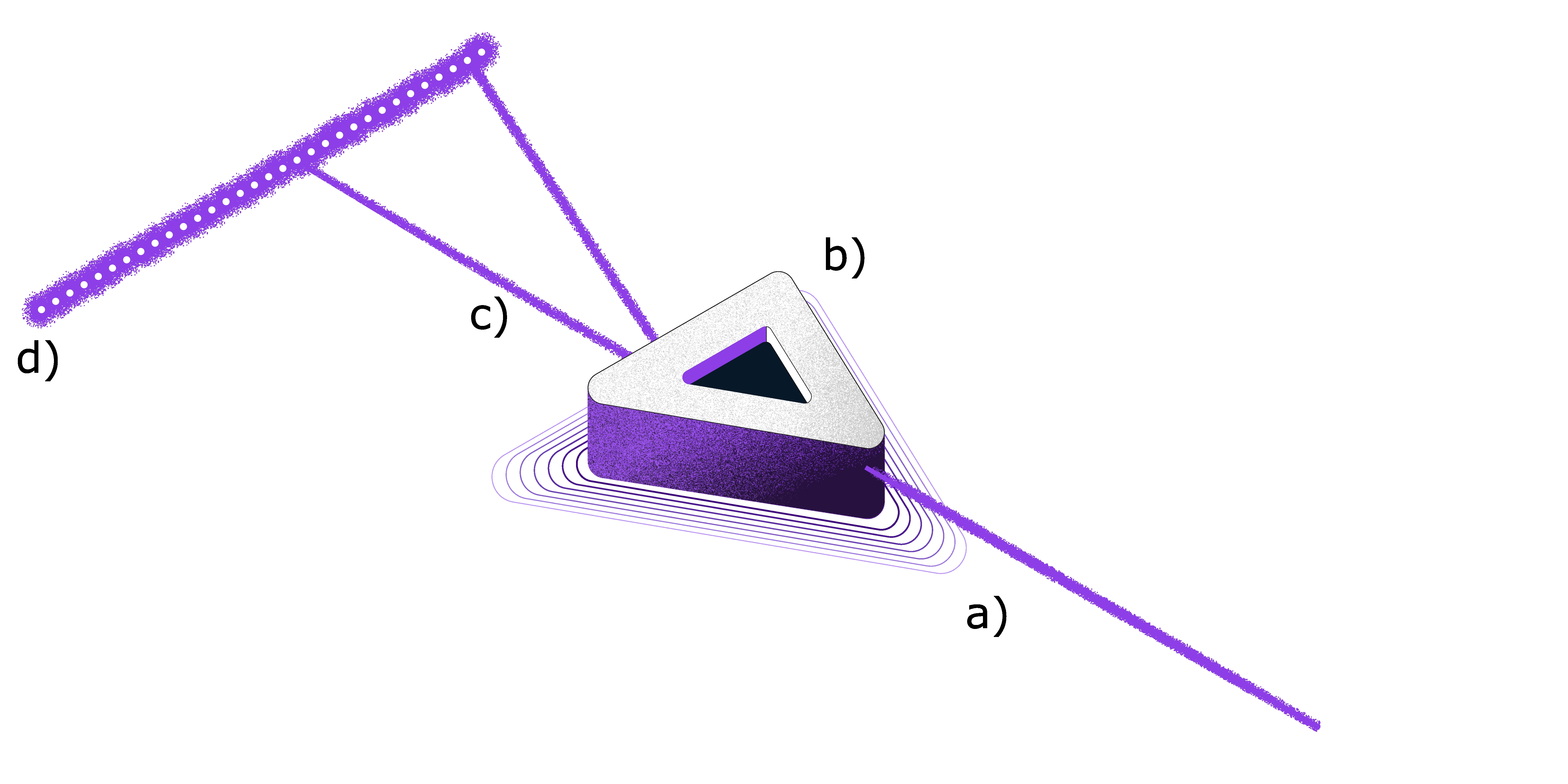}
  \caption{
  An illustration of the AOD architecture in IonQ Forte. 
  A beam (a) enters the AOD (b) and by modulating the frequency applied to the AOD device an output beam (c) can be directed onto each individual ion (d). 
  Our control system allows for alignment of beams to ions by changing the applied RF tone to the AOD. This allows each ion beam pair to be aligned with a negligible error.
  }
  \label{fig:aod}
\end{figure}

Fig.~\ref{fig:forte} depicts a schematic of IonQ Forte, which we describe next. To perform quantum gate operations, a pulsed laser at $355$~nm is used to drive the two-photon Raman transition between qubit states. 
Unlike our previously released quantum computers, which utilized multichannel acousto-optic modulators (AOM) and constrained qubit positions to align with the AOM's beam pitch, Forte is equipped with four acousto-optic deflectors (AODs) that allow for steering two counter-propagating beam pairs along the trap axis, as shown in Fig.~\ref{fig:aod}, making it possible to independently align each beam to each ion~\cite{Kim2008DopplerFree,Pogorelov2021compact}.
The capability of continuous frequency tuning in AODs reduces the beam alignment errors across the chain and also allows more flexible trapping potentials to utilize non-uniformly spaced ion chains in the system~\cite{Home_2011}. 
In addition, the laser focal size can be optimized to balance the beam pointing noise, which is stronger when using a small focal size, with the nearest neighbor cross-talk errors~\cite{Li2023lowCrosstalk}.
Using the same laser beam to address different qubits reduces the required average laser power per qubit, which allows the same optical design to access more qubits.

Before each AOD, we use an AOM to provide the necessary control of amplitude, frequency, and phase of each individual laser beam. 
Each laser beam is focused to approximately $1.5~\upmu$m waist along the ion chain axis to realize the individual addressability. 
Two of the four beam paths are designed to be capable of applying an optical phase-insensitive single-qubit gate to arbitrary qubits.
The four laser beams travel along the quantization axis of the system, set by an applied magnetic field, and are configured to form two counter-propagating beam pairs with perpendicular linear polarizations.
Each counter-propagating beam pair can be used to apply an optical phase-sensitive single-qubit gate to an arbitrary qubit by steering the beam to its location and setting the laser frequency resonant with the qubit transition.
The aforementioned two counter-propagating beam pairs enable the entanglement of arbitrary pairs in the ion chain for two-qubit gate operations. 
The current design is capable of addressing $40$ ions and $\binom{40}{2}=780$ pairs, though the present study uses the $36$-ion ($30$-qubit) configuration due to the number of photon detection channels currently available.
Throughout this work, qubit labeling or indexing corresponds  to the actual physical ion ordering in the ion trap.

An arbitrary-entangling-angle two-qubit gate, $XX_{ij}(\chi)$, is implemented by the M\o{}lmer–S\o{}rensen scheme using amplitude-modulated (AM) pulses~\cite{sorensen2000entanglement,choi2014optimal,grzesiak2020efficient,blumel2021power}, where $i$ and $j$ are qubit indexes and $\chi$ represents the angle of entanglement, and $X$ is the Pauli operator. 
Each AM profile is optimized numerically to close the phase space trajectories of all the transverse modes parallel to the trap surface for the optimal performance and the detuning frequency is chosen to be above the $20$th high frequency transverse mode to maintain insensitivity to stray fields.   
Each two-qubit gate, $XX_{ij}(\chi)$, is surrounded by two $\nicefrac{\pi}{2}$ single-qubit gates on each side using the same counter-propagating beam pairs for $XX_{ij}(\chi)$ to form a phase-insensitive two-qubit gate, $ZZ_{ij}(\chi)$~\cite{lee2005phase}. 
The total duration of each single-qubit gate is 110~$\upmu$s, which comprises nine $\nicefrac{\pi}{2}$ pulses according to the SK1 dynamical correction sequence~\cite{brown_2004CompositePulse}. 
The duration of $ZZ_{ij}(\chi)$ depends on the qubit pair we intend to entangle, and the average duration is about 900~$\upmu$s.

The quantum computer is managed by a software run-time that automates calibration of native gates and execution of circuits. 
Circuits in the native SK1 and $ZZ_{ij}$ representation are executed by an embedded control system that applies a sequence of RF pulses to the AODs and AOMs to perform coherent steering and modulation of the Raman strength. 
The software run-time maintains optimal pulse parameters through check-and-update cycles interspersed between batches of circuit executions. 

 To execute an application on the quantum computer, a quantum circuit is first submitted through a cloud interface. The quantum circuit is optimized to reduce its gate count (when possible) and compiled to native gates to match the hardware architecture. Subsequently, 
 it is then downloaded to Forte for execution. Generally, the circuit qubits are remapped to physical qubits, though the component-level benchmarks presented here are in terms of physical qubits.
 
 To enable the potential use of error mitigation by symmetrization~\cite{symm2023}, each circuit is compiled into 25 \emph{variants} that differ by local gate decomposition yet result in the same measurement statistics in the absence of noise. 
 Each variant also has a different circuit-qubit to physical-qubit assignment to further diversify the accumulation of noise. 
 Measurement statistics from each variant implementation are  collected and aggregated either by simple averaging or by plurality voting into a single histogram~\cite{symm2023}.  This type of error mitigation reduces bias in noise accumulation, and is utilized when we compute the end-to-end \#AQ benchmark as described in the next section.

\section{Benchmarking}
\label{sec:bench_and_char}

\begin{figure}[tb!]
  \centering
  \includegraphics[width=\columnwidth]{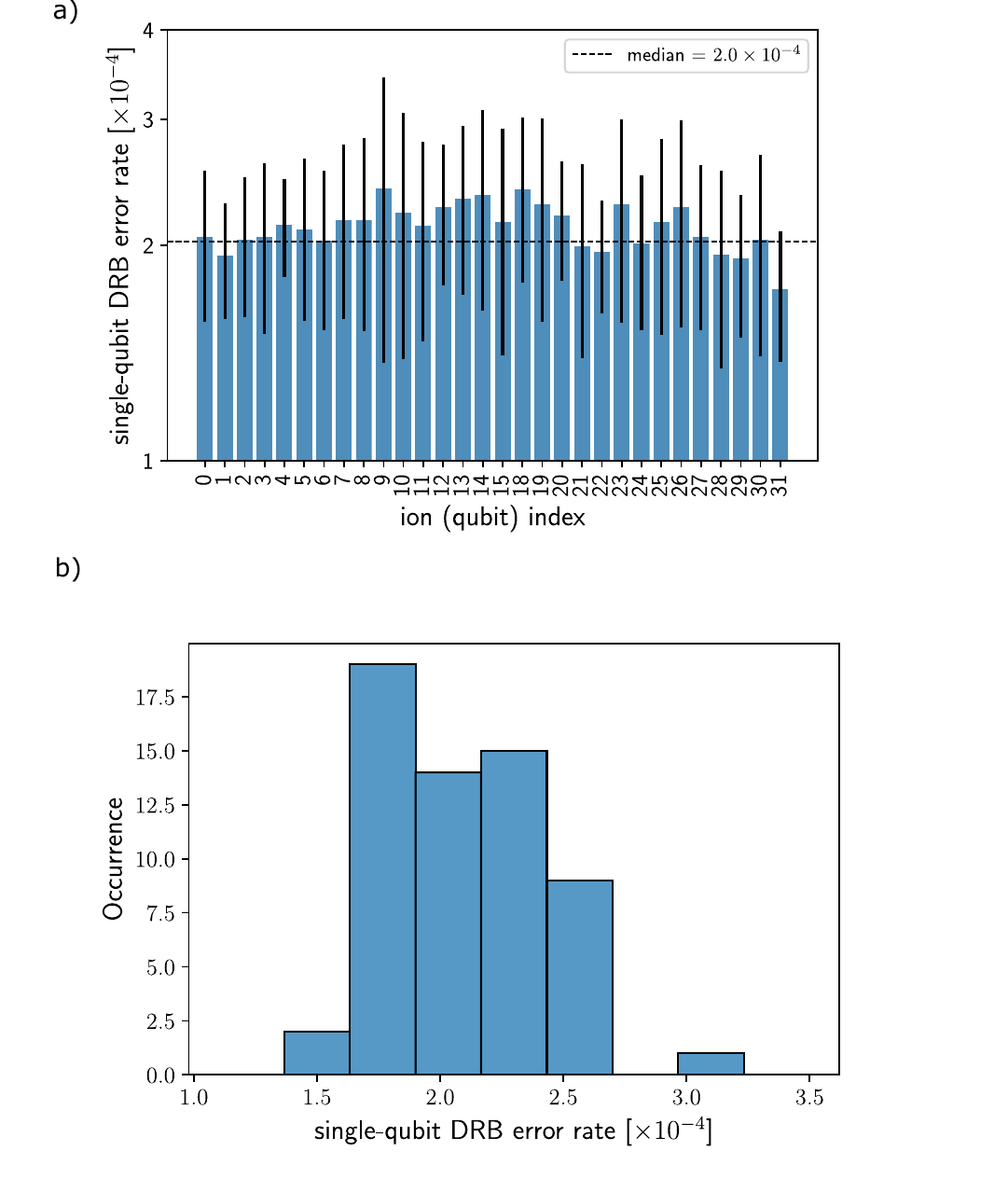}
  \caption{
  Single-qubit direct randomized benchmarking (DRB)~\cite{proctor2019drb} error-rate distributions of Forte.  
  Panel (a) shows the error rate per qubit. Error bars give standard deviation across multiple DRB runs on the same qubit, and a black dashed line indicates the median of the per-qubit and -beam-path mean values. 
  The histogram of mean per qubit and beam path error rates is shown in (b).  The quantiles in (b) are Median: $2.0\times 10^{-4}$; $10^{\mathrm{th}}$\,percentile: $1.8\times 10^{-4}$; $90^{\mathrm{th}}$\,percentile: $2.6\times 10^{-4}$.}
  \label{fig:drb_1q}
\end{figure}

\begin{figure}[tb!]
  \centering
  \includegraphics[width=\columnwidth]{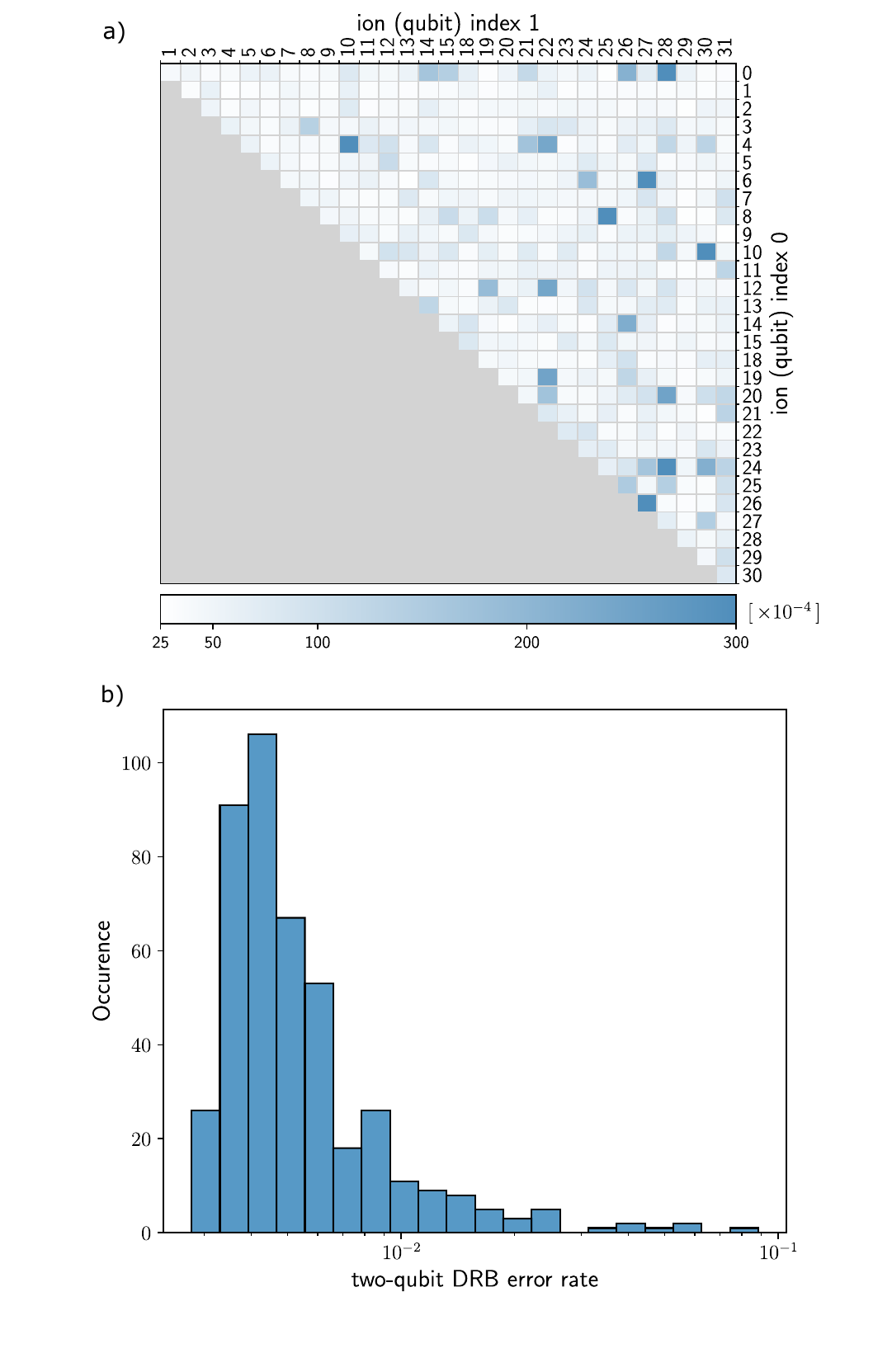}
  \caption{
   Two-qubit error rates of Forte, measured by direct randomized benchmarking~\cite{proctor2019drb}.  
   (a) Mean error rates for each qubit pair.  Seven pairs saturate the color scale with error rate $> 300\times 10^{-4}$ (see supplemental data for precise values). 
   (b) Aggregated mean error rates per pair, with Median: $46.4\times 10^{-4}$; $10^{\mathrm{th}}$~percentile: $34.5\times 10^{-4}$; $90^{\mathrm{th}}$\,percentile: $99.6\times 10^{-4}$, best observed infidelity: $27.8\times 10^{-4}$.  
   Note the logarithmic x-axis scale means a confidence interval of the DRB error rate is visually wider for smaller error rate values.}
  \label{fig:drb_2q}
\end{figure}

Benchmarks set expectations of how well quantum algorithms will perform on a quantum computer.  
Component-level benchmarks assess the performance of individual pieces (components) of a computer, and by combining these metrics the expected success of an algorithm can be estimated.   
Application-level or application-oriented benchmarks assess the performance of entire algorithms directly.  

It may seem that application-oriented benchmarks are then superior, since the process of combining component metrics is not perfect, leaving out types of and interplay between errors that do not show up when testing individual components.  
This is true when an application of interest coincides with one of the benchmarks, but this is often not the case.
Extrapolating performance from one complex algorithm to another,  based on similar circuit size for example, is possible but imperfect.
Instead, one can estimate performance based on component-level benchmarks (such as gate infidelities).  
Component-level benchmarks also give us specificity about the location of errors (\emph{e.g.}, a defective qubit), allowing us to fix problems that would be difficult to track down from application-level benchmarks alone.
Both types of benchmarks are valuable and contribute to understanding a QPU's performance.  
We consider both types separately in the following sections, and then turn to the question of reconciling their predictions.

\begin{figure}[h]
  \centering
  \includegraphics[width=\columnwidth]{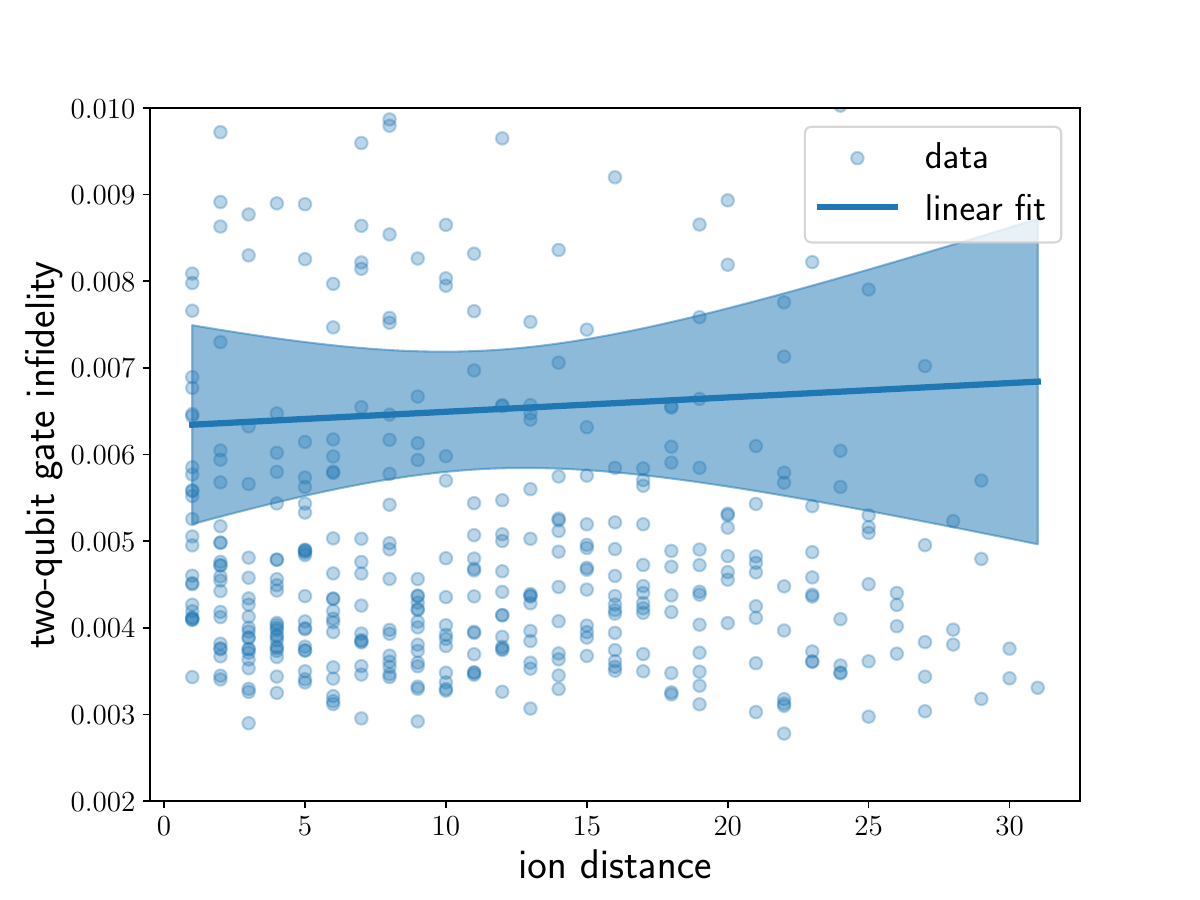}
  \caption{
  Two-qubit DRB error rate with respect to the ion distance in units of the neighboring ion spacing $3~\upmu$m. A linear model is fit to the data to estimate the correlation and the shaded region represents $2\sigma$ confidence of fit. Although the plot only shows infidelities less than $100\times 10^{-4}$, all the data are included in the fitting. The slope $0.17(45)$ indicates no statistically significant correlation between infidelity and ion distance.}
  \label{fig:2q_drb_vs_distance}
\end{figure}

\subsection{Component-level benchmarking}
\label{sec:comp_bench}

We choose as components the individual single-qubit and two-qubit quantum gates that form circuits and perform direct randomized benchmarking (DRB) \cite{proctor2019drb,polloreno2023drbtheory} on them.  
The DRB protocol involves selecting circuits by randomly sampling circuit layers according to a chosen distribution.  
For each circuit depth $d$ in a chosen set of circuit depth values, $N_c$ random circuits are selected.  
Each random circuit is executed $N_s$ (a number of samples) times, and then a success probability is computed from the measured outcome frequencies.  
The decay of average success probability with respect to circuit depth is fit with an exponential decay curve and we extract an error rate measuring the quality of a (random) circuit layer, averaged according to the chosen sampling strategy.  
In situations where the noise is known or restricted to be purely stochastic, the DRB error rate is equal to the average entanglement infidelity of a circuit layer~\cite{polloreno2023drbtheory}.  
We perform DRB using its implementation in the \texttt{pygsti} software package~\cite{pygstiPaper}.

DRB on each single qubit is performed using parameters $N_c=4$, $d \in \{1, 10, 100, 1000\}$, and $N_s=100$.  
The gate set is taken to consist of $x$- and $y$-axis $\pi/2$-rotation gates which occur with equal probability.  DRB is run on each qubit at least four times, and Fig.~\ref{fig:drb_1q} presents the obtained values.  Fig.~\ref{fig:drb_1q}a plots the error rates for each qubit, with error bars indicating the standard deviation (spread) the $\ge 4$ repetitions of DRB on the same ion.  
The histogram of mean per-qubit rates in Fig.~\ref{fig:drb_1q}b shows that the error rates are nicely centered about their median value of $2.0\times 10^{-4}$.
The raw DRB decay curves for several selected ions are given in Appendix \ref{sec:appendix_drb}.

Benchmarking a pair of qubits performs two instances of DRB, each with parameters $N_c=4$, $d \in \{1, 5, 22, 100\}$, and $N_s=100$.  
Random circuit layers are chosen by selecting the two-qubit gate ($XX_{ij}$) with probability $p_{2Q}$ or a random pair of single qubit gates with probability $(1 - p_{2Q})$.  
The two instances correspond to $p_{2Q} = 0.25$ and $0.75$, respectively.  
This allows us to extract from the two decay rates, using simple linear algebra, independent error rates for the 2-qubit and 1-qubit layers.
DRB is run on all the pairs and each pair is measured at least 4 times over the span of approximately six months. 
The mean 2-qubit DRB error rate for each pair is shown in Fig.~\ref{fig:drb_2q}a.  
While most pairs are in the $35$--$100\times 10^{-4}$ range, there are a number of outliers that result in the long tail of the complete histogram (Fig.~\ref{fig:drb_2q}b). 
The median of this distribution is $46.4\times 10^{-4}$, and its tail reaches up to $885\times 10^{-4}$ for the worst pair.
Given the large number of pairs and finite gate speed, the above values of $N_c$, $d$, and $N_s$ are chosen to strike a balance between precision and run time. Error bars on these 2-qubit error rates are 20-40\% of their reported values.
This is adequate for our purposes, though more precise results could be obtained by increasing the number and depth of the DRB circuits used.
Appendix \ref{sec:appendix_drb} shows typical 2-qubit DRB decay curves and presents results for 2-qubit DRB run with circuits with depth 1000.

As the number of ions increases to be large in a single chain, it has been shown that the entangling angle for fixed duration will scale as $1/r^3$, where $r$ is the distance between the two ions in the gate \cite{Landsman2019longchain}.
Similarly, the optimal gate duration should scale with ion distance as spectral crowding of the transverse modes starts to dominate infidelity \cite{Wu2018_NoiseAnalysis}. 
In Fig.~\ref{fig:2q_drb_vs_distance}, we plot the 2Q DRB error rate versus ion distance and find no statistically meaningful correlation between them.
This indicates that for IonQ Forte (with the above long chain) these effects do not play a role, \emph{i.e.} gate infidelity is not limited, at this time, by chain length.

\subsection{Application-oriented benchmarking}
\label{app_bench}

Ultimately a quantum processor will be used to run practical applications.  Exactly what circuits constitute ``practical applications'' has no concrete answer, and will vary from user to user.  
The performance of classical computers is measured by running representative applications collected into benchmarking suites.  Similarly, a representative set of quantum circuits can serve as a proxy for broad classes of quantum algorithms users may run.  
Quantum applications of real practical value require resources beyond what today's quantum processors can offer, and so application-oriented benchmarks use downsized versions of such applications.  
A good application-oriented benchmark will run circuits that are smaller versions of practical applications that preserve the essence of the algorithms utilized.

In this work, we run circuits from a modified version of the application-oriented benchmark set forth by the QED-C collaboration \cite{qedcPeerReviewed}.  It includes circuits for six classes of algorithm, corresponding to six practical applications 
(Hamiltonian simulation, phase estimation, quantum Fourier transform, amplitude estimation, variational quantum eigensolver (VQE) simulation, and Monte Carlo sampling). 
It omits circuits for the Bernstein-Vazirani, Deutsch-Jozsa, hidden shift, Grover's search, and Shor's algorithms because all their instances are either very shallow (all but Shor's) or very deep (Shor's).\footnote{Bernstein-Vazirani, Deutsch-Jozsa, hidden shift are strictly easier to run than phase estimation; Grover's search compiles to a very shallow circuit; the smallest instance of Shor's is $> 1000$ 2-qubit gates deep.}

For each class of algorithm, circuits are run for a range of widths (number of qubits).
In order to more fully probe the input and output space of an algorithm there are multiple circuits for each width in all algorithm classes except Hamiltonian simulation and Monte Carlo sampling.
A complete list the selected circuits can be found in the supplemental data.

\begin{figure}[tb!]
  \centering
  \includegraphics[width=0.9\columnwidth]{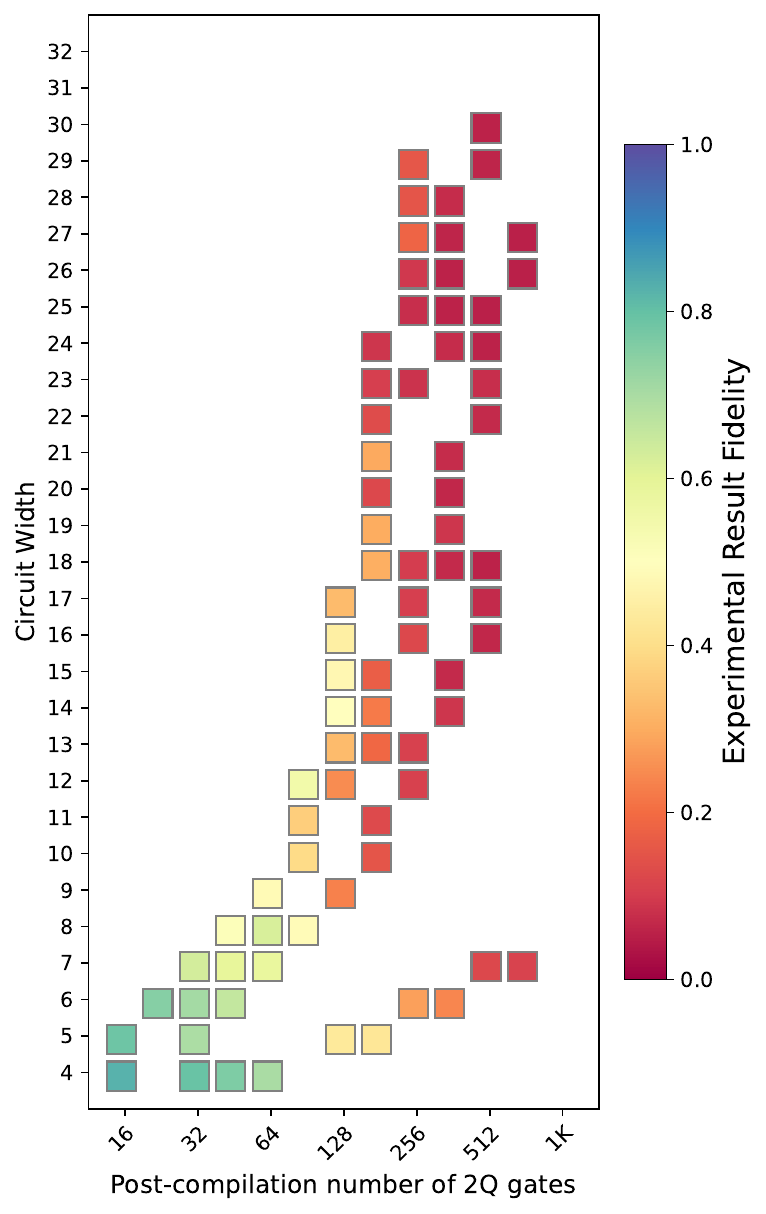}
  \caption{
  Application circuit fidelity on Forte using simple aggregation.  The x axis is the 2-qubit gate counts of the quantum circuits after all device-specific compilation, and the y axis is the circuit width (number of qubits) involved in the quantum algorithms. Colored squares indicate the average fidelity among the application instances with a given width and range of 2-qubit gate counts.  Fidelity (Eq.~\ref{eq:HellingerFidelity}) is computed between the ideal and simply aggregated (see text) data.}
  \label{fig:aq29_noem}
\end{figure}

Circuits are generated for each application instance using the procedures specified in the QED-C repository~\cite{qedcGithubRepository}.   
This circuit is then compiled into 25 variant circuits, each in terms of the available native gates.  The barriers present in the QED-C circuits are respected at all levels of compilation up to local gate optimization.
The purpose of multiple variants (e.g., instead of a single best-compilation) is to avoid the buildup of systematic errors, such that when data from these variants is aggregated together errors will be randomized.
Each variant circuit implements the same unitary but using a different physical circuit.  For example, the circuit-qubit to physical-qubit mapping may change between variants or gate bases may be randomized.  
The variant circuits are run using $30$ physical qubits, and each is executed 100 times. 

The fidelity of circuit $c$ is defined as~\cite{qedcPeerReviewed}
\begin{equation}
    F_c(p,q) = \left( \sum_i \sqrt{p_i q_i} \right)^2,\label{eq:HellingerFidelity}
\end{equation}
where $p$ and $q$ are the measured and ideal outcome probability distributions of $c$, and $i$ indexes the outcome bitstring.  
$F_c$ ranges between $0$ and $1$ and indicates the degree of overlap between two probability distributions.
(It is equal to $(1 - H^2)^2$, where $H$ is the Hellinger distance between the probability distributions.)
Eq.~\ref{eq:HellingerFidelity} differs from the final fidelity suggested in \cite{qedcPeerReviewed} by omitting a normalization to the uniform distribution.
Such normalization is insignificant for wide circuits with concentrated output distributions, and omitting it simplifies the fidelity metric.

Figure \ref{fig:aq29_noem} shows the average value of $F_c(p,q)$ for circuits of a given size.  The outcomes of the 25 variant circuits are simply added together (after permuting the qubit labels if needed) to form an overall outcome histogram containing 2500 counts for each algorithm instance.
We refer to this process as the \emph{simple aggregation} of variant circuits histograms to distinguish from a more complicated plurality-vote method later on.
Fidelity (Eq.~\ref{eq:HellingerFidelity}) is computed for each application instance using the normalized histogram, and colored boxes in Fig.~\ref{fig:aq29_noem} indicate the average fidelity over all the application instances of a given size. 
This volumetric plot indicates that fidelity degrades smoothly as circuit width and depth increase, as we might expect from a simple error model.  The fidelity remains above $1/e \approx 0.37$ for circuits with widths up to $\approx 20$ qubits and depths up to $\approx 200$ two-qubit gates.

\begin{figure}[tb!]
  \centering
  \includegraphics[width=\columnwidth]{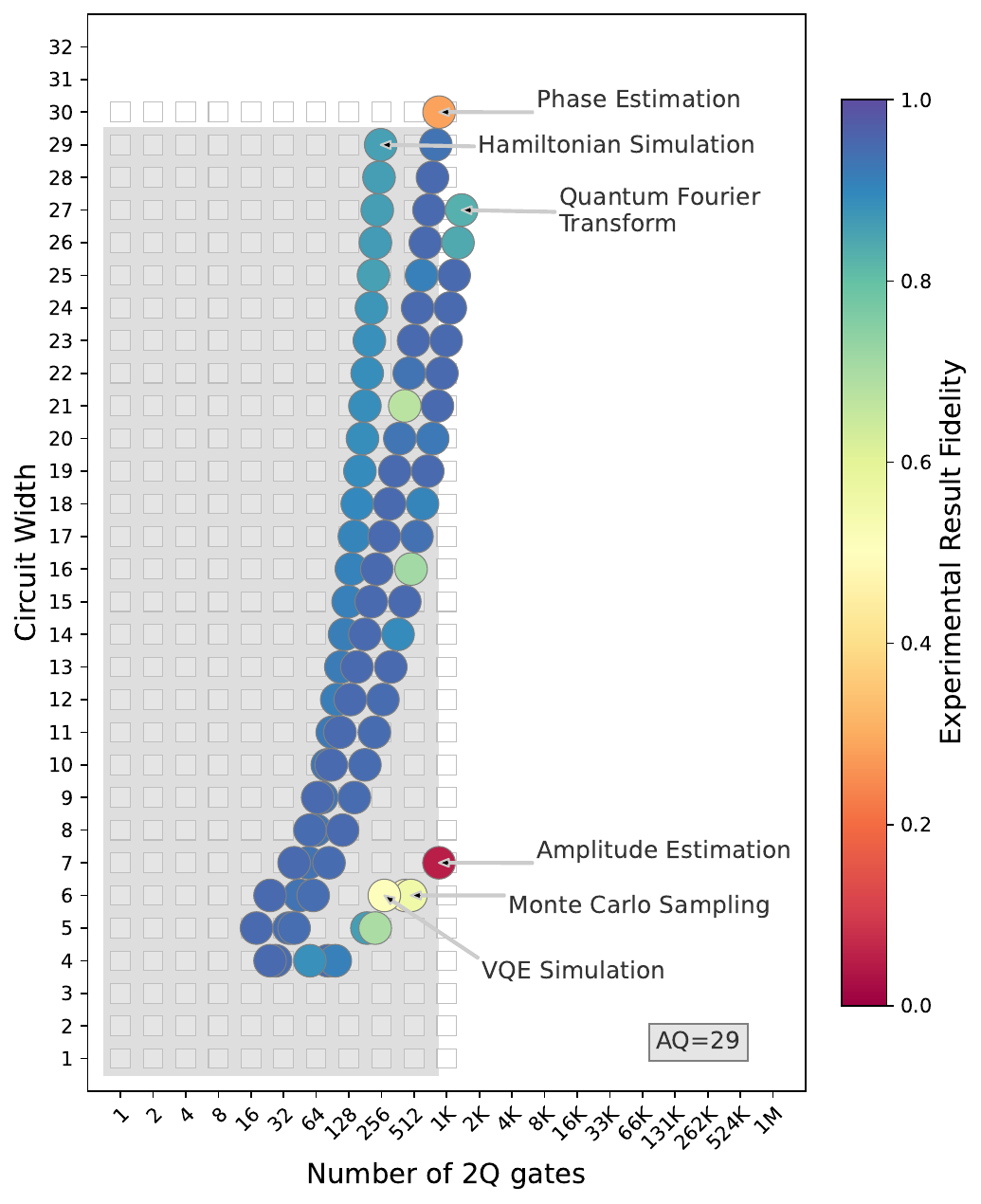}
  \caption{
  Application circuit fidelity results showing 29 algorithmic qubits on Forte.  Axes indicate the 2-qubit gate count of quantum circuits prior to device-specific compilation ($d_c$) and the the circuit width ($w_c$) respectively. Colored circles indicate the minimum fidelity measured for each (algorithm, $d_c$, $w_c$) triple.  A plurality-vote method is used to aggregate observed outcomes and fidelity (Eq.~\ref{eq:HellingerFidelity}) is computed between this error-mitigated distribution and the ideal.   Color scale is identical to Fig.~\ref{fig:aq29_noem} to facilitate comparison.  A gray shaded region indicates where circuit fidelities meet the \#AQ criterion, and corresponds to a \#AQ value of 29.}
  \label{fig:aq29}
\end{figure}

Our choice of circuits was motivated by an end-to-end performance benchmark called the number of algorithmic qubits (\#AQ).
This volumetric benchmark \cite{BlumeKohout2020volumetricframework} summarizes quantum computer performance in a single number called the ``number of algorithmic qubits (\#AQ)''.
It is based on the data used to generate Fig.~\ref{fig:aq29_noem} but,
because it is intended to measure end-to-end performance, the \#AQ benchmark incorporates the pre- and post-processing gains due to compiler optimizations and error mitigation.  

Compilation is included by treating the initial QED-C-generated circuits as reference implementations of each application instance.  These reference circuits provide a common starting point for all users of the benchmark.  
The width and number of two-qubit gates in each reference circuit $c$ are denoted by $w_c$ and $d_c$, respectively.

Error mitigation refers to a classical post-processing step that attempts to amplify the signal and minimize the noise in the outputs from the quantum computer.  In our case we utilize a plurality-vote procedure \cite{symm2023} that constructs a final outcome histogram from the outcome histograms of the variant circuits in a simple but non-linear way (see Appendix \ref{sec:appendix_pluralityvote} for details on this procedure), instead of the standard addition-of-histograms aggregation leading to Fig.~\ref{fig:aq29_noem}.

A quantum computer has (at least) $n$ algorithmic qubits when all the application circuits (for all applications) with $w_c \le n$ and $d_c \le n^2$ run with fidelity greater than $1/e$.  The largest such $n$ defines the \#AQ for the computer.  
Note how the end-to-end nature of the benchmark is visible in the above definition: $w_c$ and $d_c$ relate to the size of the \emph{reference} circuit prior to device-specific compilation, and that the fidelity is taken between the ideal distribution and the \emph{error-mitigated} outcome distribution.

Figure \ref{fig:aq29} shows the volumetric plot used to determine the device's \#AQ value.
Its axes indicate the reference circuit $d_c$ and $w_c$, and fidelity values are computed using error-mitigated outcome distributions.

Forte achieves \#AQ=29, which means that all the circuits falling within the shaded rectangle in Fig.~\ref{fig:aq29} (covering the region $w_c \le 29$, $d_c \le 29^2 = 841$) pass the success threshold test stated above. 
For the work presented here, we are limited to \#AQ29 by the amplitude estimation class of circuits at circuit width 7 (AE7), which contain on average 604.25 2-qubit gates after compilation and have an average post-error-mitigation fidelity of 0.125.  30-qubit wide phase estimation circuits, on the other hand, have on average 471 2-qubit gates and an average post-error-mitigation fidelity of 0.78.  This follows the generally observed feature that \#AQ is depth- rather than width-limited in our system. 
Comparing Figs.~\ref{fig:aq29_noem} and \ref{fig:aq29}, which have the same color scale, we can see how effective the error mitigation is at reducing noise in the circuit outcome counts.
The plurality vote procedure is particularly effective at eliminating noise in outcome histograms whose ideal distributions have weight concentrated on just one or several bit strings.
This property is present for five out of the six circuit families (all but VQE) included in the \#AQ benchmark.

\begin{figure}[tb]
  \centering
  \includegraphics[width=\columnwidth]{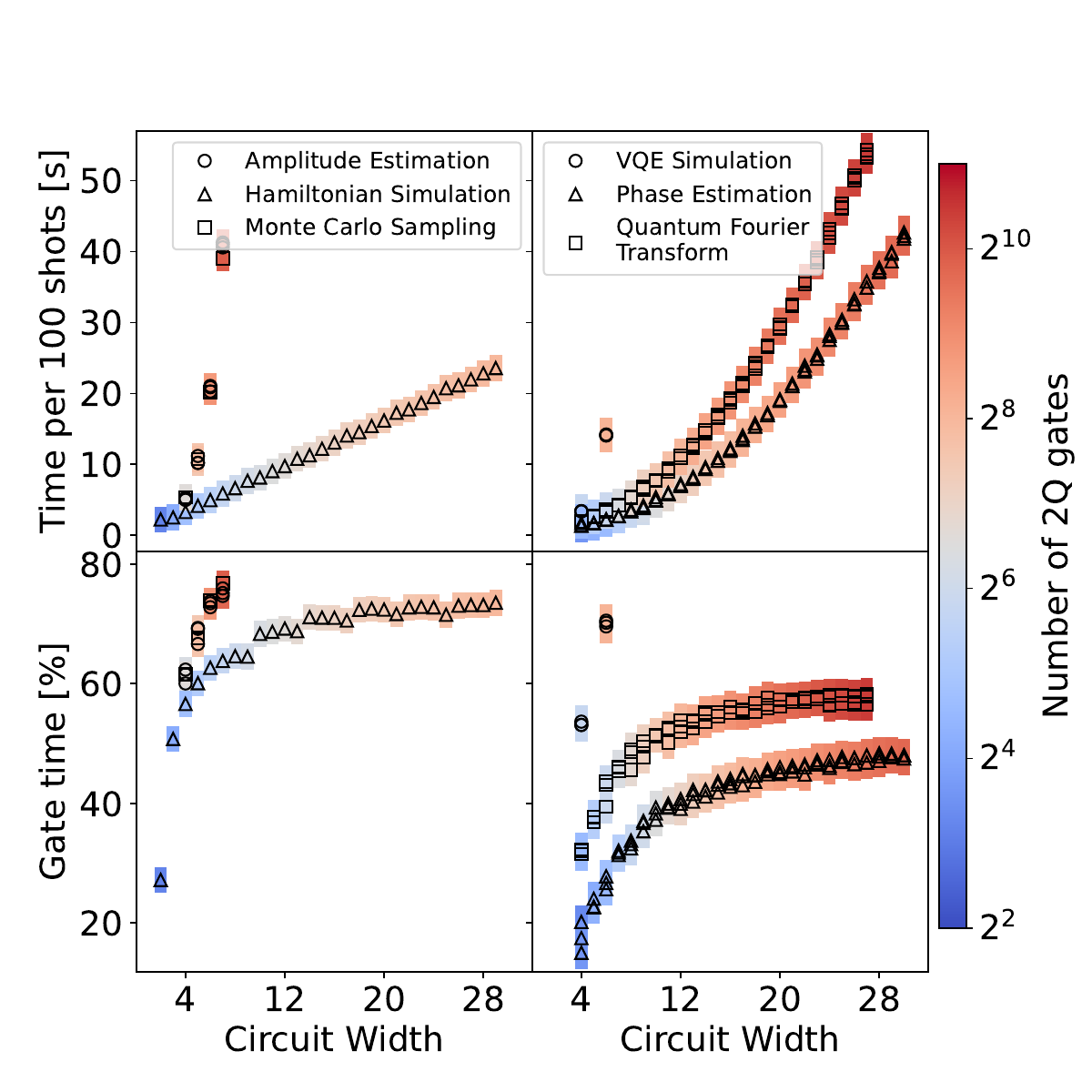}
  \caption{Run timing performance metrics for application-oriented benchmark circuits. The top row shows the on-board execution time for 100 shots of each circuit, which \emph{includes} communication between the cloud and the quantum computer, ion cooling, state preparation, quantum circuit execution, state detection, and shot repetition latency. The bottom row shows the fraction of that time actually spent on quantum gates and AOM, AOD switching. Note, overheads from system calibration, circuit compilation, and result processing for error mitigation are  excluded.
}
  \label{fig:execution_time}
\end{figure}

The execution time of each variant, which is defined as the duration between when the cloud submits a circuit to Forte and when the cloud receives the computation result from Forte, is summarized in Fig.~\ref{fig:execution_time}. This includes classical overhead, such as communication between the cloud and the quantum computer and shot repetition latency; as well as quantum overhead, such as quantum state preparation and detection, but excludes system calibration, circuit compilation, and data processing occurring in the cloud. The gate time percentage in Fig.~\ref{fig:execution_time}, which only considers the time spent on the quantum circuit execution, is calculated using the average two-qubit gate time and between-gate padding time that accounts for switching of acousto-optic devices.

\section{Simulation of application circuits}
\label{sec:sims}

We are interested in whether our set of component-level benchmarks is consistent with the performed application-oriented benchmarks.  
The general question of whether component-level benchmarks can be used to predict application performance has significant practical ramifications, since running an application-level benchmark is 1) less general, in that it targets only a subset of all possible applications and 2) typically much more resource intensive than component-level benchmarks (as they typically consist of deeper and wider circuits than component-level benchmarks).  This second point is especially true as quantum processor (and thereby application) size increases.  

If component-level benchmarks \emph{do} predict application performance, this would not only result in considerably less resources being required to assess application performance, but also suggest that the types of errors not included by the modeling that maps component- to application-level performance are insignificant.  These unmodeled errors often include many of the most insidious types of errors such as crosstalk and context dependence.

We investigate the extent to which a depolarizing noise model based on measured DRB error rates (Sec.~\ref{sec:comp_bench}) can explain the performance of algorithmic qubit circuits measured in \ref{app_bench}.
By using such a simple noise model we ignore structure in the errors even at the component level.  This is supported by 1) results from gate set tomography (GST)~\cite{nielsen2021gate} on just a few qubit pairs that indicate the dominant errors are stochastic in nature, and 2) our goal being to investigate whether there is a major source of error not accounted for by the individual components, \emph{e.g.}, drift and/or crosstalk errors.  For this purpose, a depolarizing noise model will suffice, and we defer an error analysis using more detailed error models to future work.

\begin{figure}[tb!]
  \centering
  \includegraphics[width=\columnwidth]{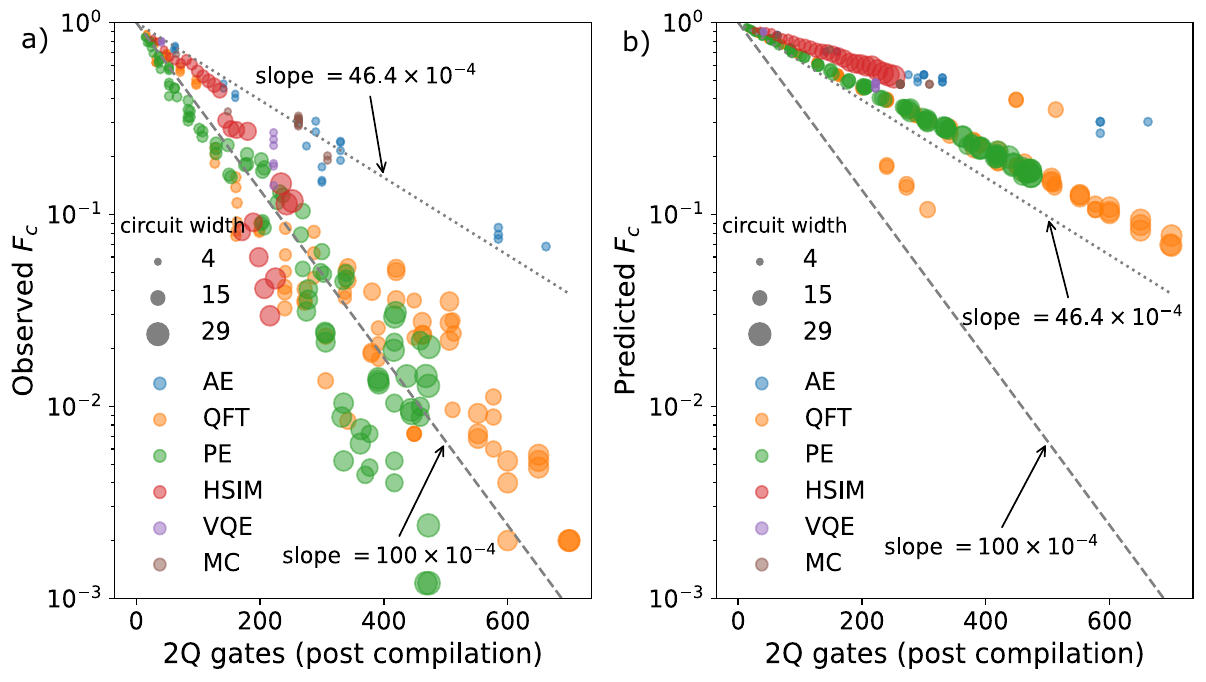}
  \caption{Observed and simulated fidelity of circuits in an application-oriented benchmark without error mitigation.  (a) Fidelity (Eq.~\ref{eq:HellingerFidelity}) between the observed and ideal outcome distributions, the former computed via simple aggregation of variant-circuit histograms.  Each point marks the fidelity of a single application instance.  Colors indicate algorithm class and point sizes circuit widths.  
  (b) Simulated results for the same circuits using a depolarizing noise model with $\epsilon_{1Q} = 2.0\times 10^{-4}$ and $\epsilon_{2Q} = 46.4\times 10^{-4}$.
  Lines with slopes equal to $46.4\times 10^{-4}$ and $100\times 10^{-4}$ are shown to guide the eye.
  }
  \label{fig:aq_modeling_noem_vs_n2Q}
\end{figure}

\begin{figure}[tb!]
  \includegraphics[width=\columnwidth]{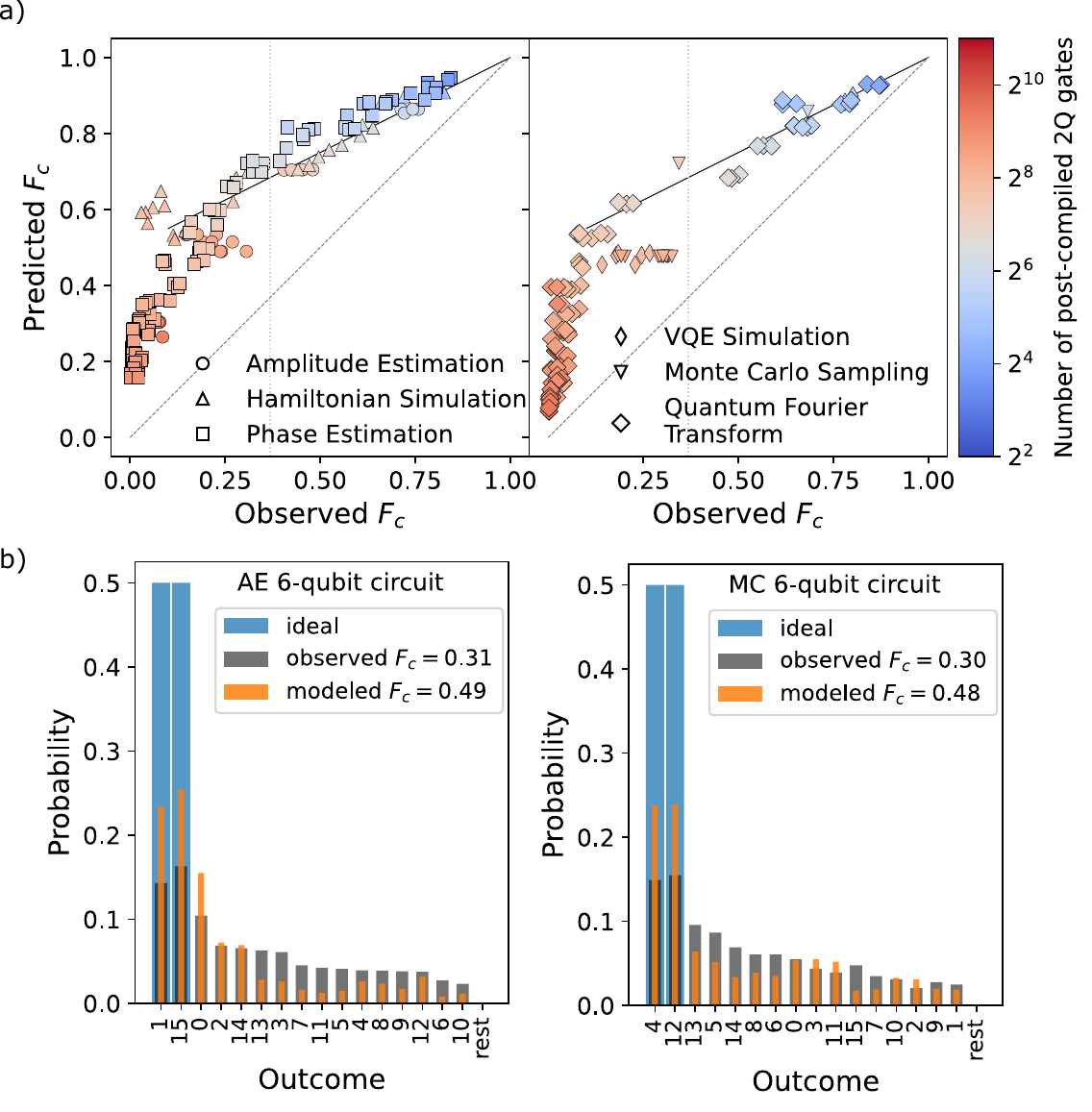}
  \caption{Comparison between simulation and observed results in an application-oriented benchmark without error mitigation. (a) Fidelity (Eq.~\ref{eq:HellingerFidelity}) between predicted and ideal vs. observed and ideal circuit outcomes for each of the application circuits executed.  Simulations use a depolarizing noise model described in the text.  Lines indicate slope $1$ (dashed), slope $0.5$ (solid), and observed $F_c = 1/e$ (dotted).
 (b) Example outcome distribution comparisons for a typical instance of an Amplitude Estimation algorithm (left) and a Monte Carlo Sampling algorithm (right) on 6 qubits. 
 Bars for the 16 largest outcomes are shown, and remaining outcomes are lumped into a single ``rest'' entry.}
  \label{fig:aq_modeling_noem}
\end{figure}

We construct the depolarizing noise model where all single-qubit gates are identical, all two-qubit gates are identical, and the error on each is given by a single depolarization rate $\epsilon_{1Q}$ and $\epsilon_{2Q}$ respectively.   A depolarization rate $\epsilon$ on $n$-qubit gates means that noisy gates are modeled as the ideal gate followed by the ($n$-qubit) channel
\begin{equation}
\Lambda_\epsilon : \rho \rightarrow (1-\epsilon)\rho + \frac{\epsilon}{4^n - 1}\sum_{P \ne I} P \rho P,
\end{equation}
where $\rho$ is a density matrix, and the sum is over all non-identity tensor products of $n$ Pauli operations. This definition implies that $\epsilon_{1Q}$ equals the entanglement infidelity of modeled single-qubit gates and $\epsilon_{2Q}$ equals the entanglement infidelity of the the M\o{}lmer–S\o{}rensen gate.  We set $\epsilon_{1Q} = 2.0\times 10^{-4}$ and $\epsilon_{2Q} = 46.4\times 10^{-4}$, the median values of the single- and two-qubit DRB error rate distributions, respectively (see Figs.~\ref{fig:drb_1q}b and \ref{fig:drb_2q}b).

Using this depolarization model, we simulate each of the variant circuits  (the final form of each circuit as run on the hardware).
Circuits are simulated using Google's Qsim simulator on a cluster of GPUs with NVIDIA's cuStateVec back-end library for efficient simulation of the full state vector \cite{IsakovQSim2021}.

Simulated outcomes are compared to observed outcomes by computing the fidelity (Eq.~\ref{eq:HellingerFidelity}) between each of these distributions and the ideal outcome distribution.  
We combine variant circuit results using simple aggregation rather than plurality-voting because the model, being based on component-level benchmarks, is intended to predict the non-error-mitigated hardware performance.
Figure \ref{fig:aq_modeling_noem_vs_n2Q} plots the observed and predicted fidelity as functions of the as-executed 2-qubit gate count.
It indicates significant discrepancy between the predicted and observed performance: the observed fidelity values suggest a rough error rate of $100\times 10^{-4}$ whereas the predictions (unsurprisingly) show a rate close to the component-level 2-qubit gate error rate, $\epsilon_{2Q} = 46.4\times 10^{-4}$. 
Fig.~\ref{fig:aq_modeling_noem}~a plots the observed and predicted fidelity directly against each other.  The model's over-estimation of circuit fidelity shows itself here as a characteristic arc above the $x=y$ line.  The slope of the data points in the upper right corners of the plots roughly corresponds to the ratio between the overall actual and predicted error rates.  This ratio is dominated by the 2-qubit error rate, which from Fig.~\ref{fig:aq_modeling_noem_vs_n2Q} we see is approximately $1/2$, and the solid lines in Fig.~\ref{fig:aq_modeling_noem}~a corroborate this.   To give a sense of what the raw outcome histograms look like, Fig.~\ref{fig:aq_modeling_noem}~b, shows a comparison between the observed, predicted, and ideal outcome distributions for two representative circuits.

\begin{figure}[h!]
  \centering
  \includegraphics[width=0.9\columnwidth]{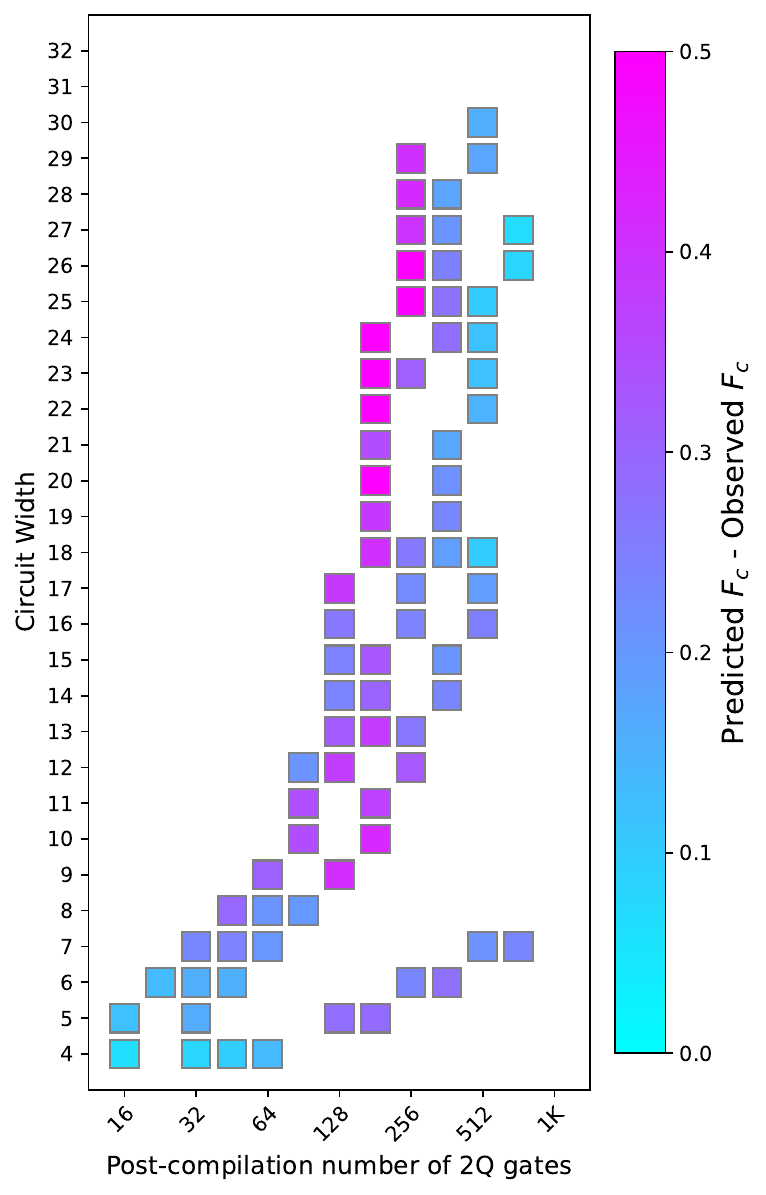}
  \caption{
  Difference between simulated and observed fidelities of application-oriented benchmark circuits.  This volumetric plot presents data in the same manner as Fig.~\ref{fig:aq29_noem}, except color indicates the difference between the values in Fig.~\ref{fig:aq29_noem} and the circuit fidelities predicted by a noise model derived from component-level characterizations.  The color scale is strictly positive since, for this data set, the predicted fidelity is \emph{always} higher than what is observed (see Fig.~\ref{fig:aq_modeling_noem} a).}
  \label{fig:aq_29_noem_sim}
\end{figure}

Lastly, in Fig.~\ref{fig:aq_29_noem_sim} we compare the predicted and observed $F_c$ values using a volumetric plot similar to Fig.~\ref{fig:aq29_noem}.  This view of the data shows that difference in fidelity does not simply scale with circuit size, and that the different structure found in circuits for different algorithms significantly impacts how well the component-level model performs.  By referencing Figs.~\ref{fig:aq29} and \ref{fig:aq_modeling_noem} we can infer that the larger Hamiltonian simulation circuits have the largest fidelity differences, even though there are comparable and even larger circuits in the considered set.

\section{Conclusion}
\label{sec:conclusion}

In this paper, we showed the operation of IonQ Forte: a trapped-ion quantum computer configured with 30 qubits in a single linear ion chain. 
By exhaustively benchmarking all one- and two-qubit gates, we showed that high-fidelity operations are possible, and are not appreciably impacted by ion separation within the chain. 
To the authors' knowledge, this work is the longest single-chain trapped ion processor demonstrated to date, showing that performance degradation via long-range interactions does not impact performance at the present scale.

In addition to component benchmarks, we showed the results of a suite of application-level benchmarks.
We show the performance of these circuits before and after error mitigation. 
Before error mitigation, we notice that the useful circuit depths depend on the algorithm being run, but we typically see ~200 two-qubit gate applications before noise reduces the fidelity to below $1/e$.
Applying error mitigation greatly suppresses the noise to the point that all circuits within the \#AQ~29 benchmark pass the $1/e$ fidelity threshold criterion.

We tracked the runtime and non-gate overhead associated with these benchmarks, demonstrating that the single-chain architecture allows for both good performance as well as rapid time-to-solution.
We ran noisy-gate simulations of application-level benchmarks using a simple model of depolarization rates set by the component-level benchmarks. 
Though we observe a correlation between simulation and experiment, these simplistic models were unable to reproduce the experimentally-measured application-level benchmark across most of the circuits in our test suite, with the predicted performance exceeding what was observed in experiment.
Our current hypothesis is that the noise we observe is stochastic spin-phase noise, which is likely mechanical in nature and is transferred to the qubits via small optical path length fluctuations. 
Due to a non-trivial power spectral density, this noise cannot be decomposed into Markovian gate-level noise, which leads to context dependency.

As researchers strive for cases of useful quantum advantage, we face new and challenging problems stemming from the number of qubits involved and their interactions, rather than their individual basic physics. 
This is not surprising; Philip Anderson noted fifty years ago that ``More is different" \cite{anderson1972more}.
Further, as quantum computer development continues, engineering them toward reductionism (in which their global behavior can be efficiently decomposed into their constituent parts) will be critical.
Toward that end, we see great utility in system-level characterization tools such as volumetric benchmarking~\cite{proctor2022scalable}, quantum volume~\cite{cross2019validating,baldwin2022re}, layer fidelity~\cite{mckay2023benchmarking}, and quantum computer capability regions~\cite{proctor2022measuring}, among others. 
When combined with system-level simulation tools to drive hardware development in this direction, the result will be quantum computers that are not only performant, but also explainable.

\section{Acknowledgements}
The authors thank Jungsang Kim, Dean Kassmann, and KC Erb for valuable discussions and commentary on the manuscript, and Charlie Baldwin for feedback on the supplemental data.

\section{Data availability}
All data presented in this paper can be obtained from the supplemental data repository accompanying this work~\cite{suppdata}. There, we also include both the pre- and post-compiled circuit variants, raw experimental results,  raw simulation results, code for performing the error mitigation, and code for generating all figures in this manuscript. 

\bibliographystyle{quantum}
\bibliography{citations}

\appendix

\section{Direct randomized benchmarking\label{sec:appendix_drb}}

\begin{figure}[h]
  \centering
  \includegraphics[width=0.5\columnwidth]{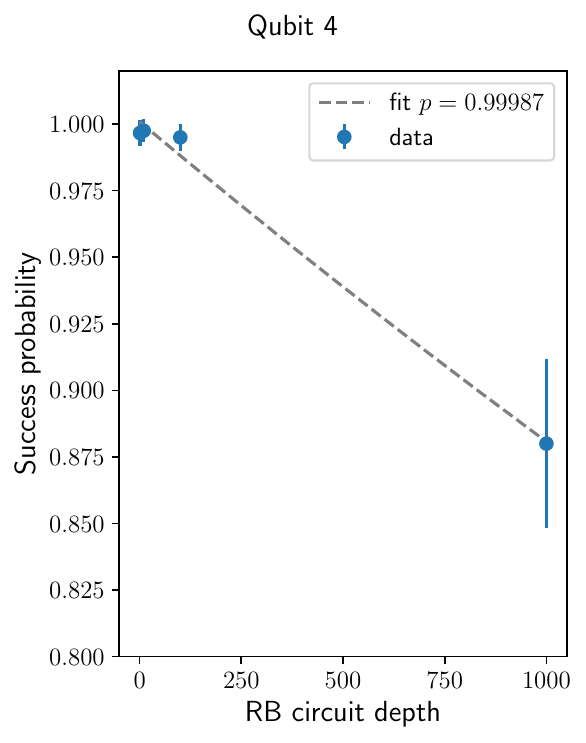}
  \includegraphics[width=0.5\columnwidth]{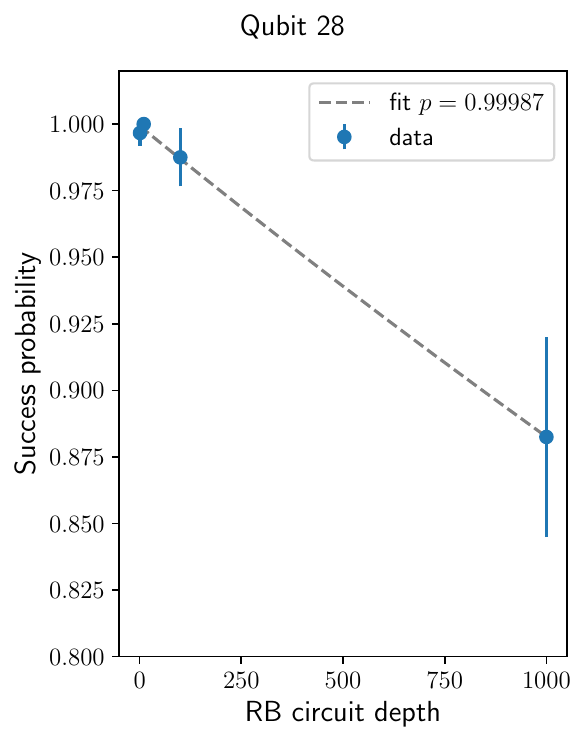}
  \caption{
  Example single-qubit DRB decay curves.  Upper and lower panes 1-qubit DRB data sets, for qubits (ions) 4 and 28, respectively.  The x-axis indicates the number of random layers used to construct the DRB circuits, and the y-axis plots the probability that the ideal outcome is observed.  Dashed line shows fit to $a + b p^d$ curve. These plots are typical results for all the ions in the chain.}
  \label{fig:drb_1q_decay}
\end{figure}

We run 1- and 2-qubit direct randomized benchmarking (DRB) \cite{proctor2019drb,polloreno2023drbtheory} to obtain the results presented in section \ref{sec:bench_and_char}.  Single-qubit DRB is performed using $N_c=4$ random circuits at each DRB depth (the number of randomly sampled DRB layers within a DRB circuit) $d \in \{1, 10, 100, 1000\}$.  Each circuit is repeated $N_s=100$ times, and ideally results in a single outcome.  The probability we observe this ``success'' outcome as a function of the DRB depth is fit to a function of the form $a + b p^d$, where $d$ is the depth and $\{a, b, p\}$ are fit parameters.  We care most about the value of $p$, the average error rate of a 1-qubit gate ($a$ and $b$ relate to state preparation and measurement errors).  A 1-qubit DRB decay curve illustrating the typical data underlying Fig.\ref{fig:drb_1q} is show in Fig.~\ref{fig:drb_1q_decay}.

\begin{figure}[h]
  \centering
  \includegraphics[width=\columnwidth]{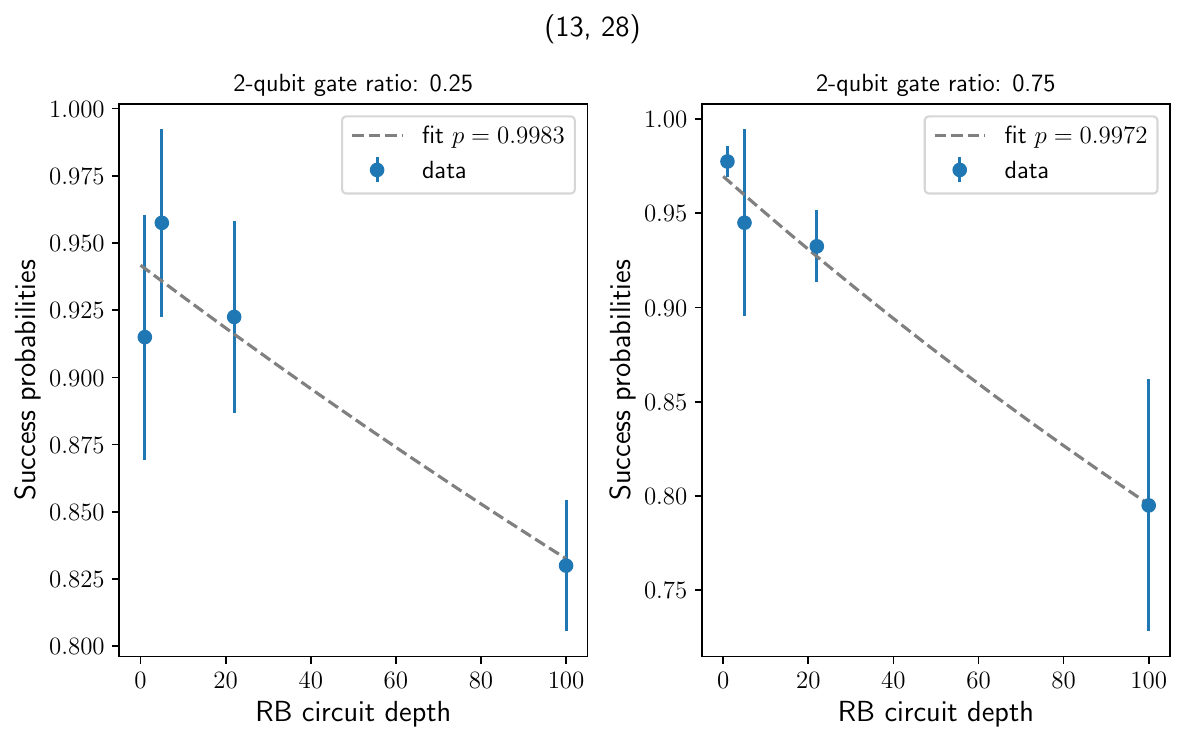}
  \caption{
  Example 2-qubit DRB decay curves.  Typical 2-qubit DRB decay curves for 2-qubit gate ratios ($p_{2Q}$) of $0.25$ (left) and $0.75$ (right).  Results here are for the qubit (ion) pair (13,28).  The x-axis indicates the number of random layers used to construct the DRB circuits, and the y-axis plots the probability that the ideal outcome is observed. Dashed line shows fit to $a + b p^d$ curve.}
  \label{fig:drb_2q_decay}
\end{figure}

To measure an average two-qubit-gate error rate we perform two instances of 2-qubit DRB that differ in the probability that a 2-qubit gate is randomly chosen ($p_{2Q}$) when constructing the random circuit layers portion of each DRB circuit.  We can then extract from the two resulting decay rates separate error rates for 1-qubit and 2-qubit gates.  To obtain the error rates presented in the main text, we collect data for $N_c=4$ random circuits at each DRB depth $d \in \{1, 5, 22, 100\}$ for $p_{2Q} = 0.25$ and $0.75$.  Each circuit is repeated $N_s=100$ times and the best-fit decay curve of the form $0.25 + b p^d$ (the same as the 1-qubit case but with $a$ fixed as $0.25$ to ensure the expected asymptote) is found.  Figure \ref{fig:drb_2q_decay} shows a set of 2-qubit DRB decay curves for a particular qubit pair, (13,28), which is typical of the dataset.  Thus, if $p_0$ and $p_1$ are the decay rates (values of $p$ from our fit) corresponding to $p_{2Q} = 0.25$ and $0.75$, respectively, then
\begin{equation}
    \begin{bmatrix} 1-\left(1-r_{1Q}\right)^2 \\r_{2Q} \end{bmatrix} = \begin{bmatrix} 0.75&0.25\\0.75&0.25 \end{bmatrix}^{-1} \begin{bmatrix} p_0\\p_1 \end{bmatrix}
\end{equation}
gives the 1- and 2-qubit error rates, $r_{1Q}$ and $r_{2Q}$, obtained from 2-qubit DRB.  We're primarily interested in the latter, as 1-qubit DRB gives us a more accurate measure the the single-qubit error rate (because it uses deeper circuits).

\begin{figure}[h]
  \centering
  \includegraphics[width=\columnwidth]{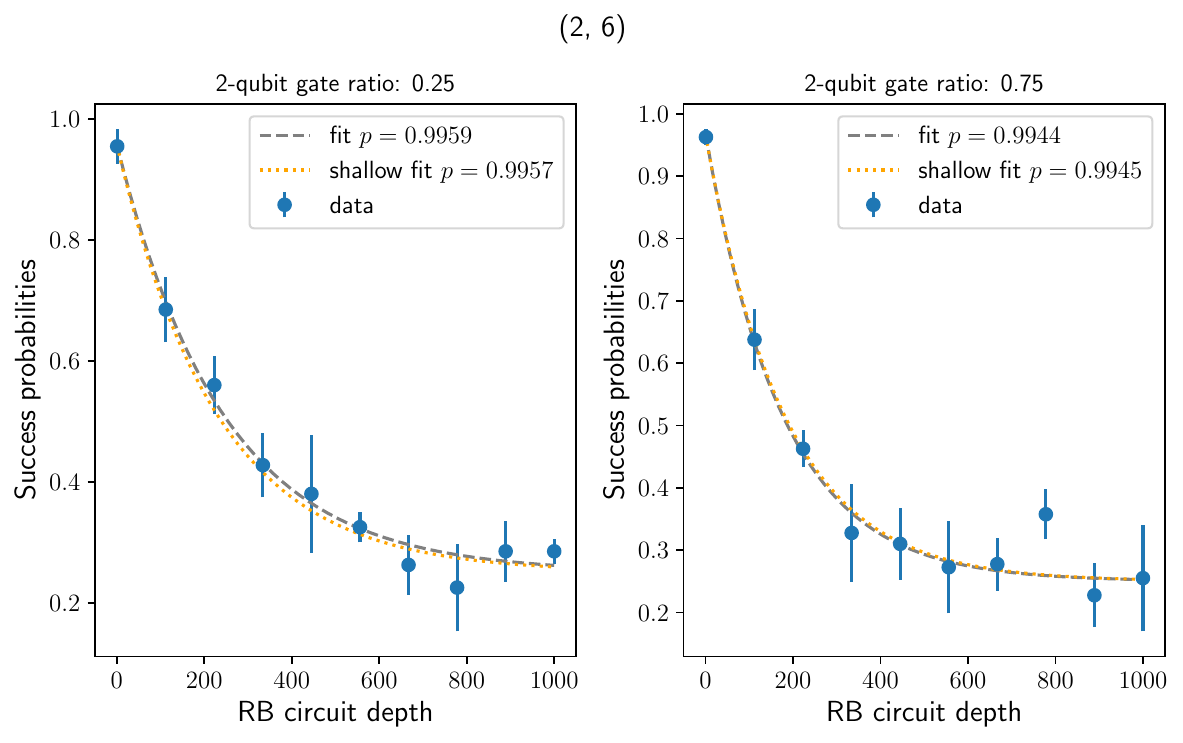}
  \caption{
    Example 2-qubit DRB decay curves with deeper circuits.  2-qubit DRB decay curves for 2-qubit gate ratios ($p_{2Q}$) of $0.25$ (left) and $0.75$ (right) and maximum depth $1000$.  Results here are for the qubit (ion) pair (2,6), and are typical. 
    Dashed gray line shows fit to $a + b p^d$ curve all the data.  Dotted orange line shows a similar fit using just the first two points (depth $\le 112$).}
  \label{fig:drb_2q_decay_deep}
\end{figure}

The choice of $N_c=4$ and a maximum depth of $100$ where made to balance run time and precision.  Parametric bootstrapping with 200 samples gives error bars that are 20-40\% of the reported 2-qubit error rates, which is adequate for our purposes.  We also performed much deeper 2-qubit DRB experiments on select pairs, using $N_c=4$ circuits at each depth $d \in \{1, 112, 223, 334, 445, 556, 667, 778, 889, 1000\}$.  Figure \ref{fig:drb_2q_decay_deep} shows a typical case of "deep" DRB decay curves.  This figure compares the full analysis of this data (dashed gray line) with one that truncates the data after the second data point to be similar to (actually sparser than) the DRB datasets we obtain using our standard DRB parameters where $d \in \{1, 112\}$ (dotted orange line).  We find that the difference between the fits' $p$-parameter values is less than 10\%, and is consistent with our bootstrapped error bar estimates for the overall error rate.  Finally, we note that the good quality of the exponential fit in \ref{fig:drb_2q_decay_deep} indicates that a single, time-independent, depolarizing channel is sufficient to explain the data.  This would not be the case if the 2-qubit gate's fidelity became significantly worse over time, for example.

\section{Plurality voting procedure\label{sec:appendix_pluralityvote}}
In this appendix we describe the plurality voting procedure used in this work to aggregate the variant histograms of an algorithm instance into a single histogram.
Plurality voting was introduced in Ref.~\cite{symm2023}, and comprises two steps. First, the total number of shots is divided into a number of \emph{variants}. Each variant is implemented in a different way (\emph{e.g.}, different qubit assignments or compilation strategy), but in the absence of noise all variants represent equivalent circuits.
Second, the histograms from each variant are post-processed by \emph{voting} as follows.
\begin{enumerate}
    \item Choose a threshold $t$
    \item Randomize the shots within each variant, arranging them in an array with columns corresponding the variants and rows corresponding to shots.
    \item For each row, determine the plurality bit string. If the number of occurrences is greater than the threshold $t$, then add it to the aggregated histogram.
    \item Re-randomize and repeat until the aggregated histogram converges. 
    \item If no bit strings have been accepted, decrease $t$ by one.
    \item If $t=2$, then stop and instead average the counts across shots to produce an aggregate histogram. This corresponds to determining that there not sufficient consensus across variants for plurality voting to be employed.
\end{enumerate}
In the present work, we use 25 variants, 100 shots each, and a starting threshold of $t=7$ for all experiments.
These parameters were chosen heuristically via initial testing on the system.

Brute-force sampling in the above aggregation scheme is not efficient, since for sufficiently high thresholds obtaining a 
plurality is a rare event.
To compute this efficiently, we note that the probability of a bit string in the output distribution is the probability of it being chosen on at list $t$ variants while no other bit string is simultaneously chosen on more variants.

The first simplification that we employ is by noting that, for a given positive integer threshold $t$, if a bit string does not appear within at least $t$ variants, it will never be chosen in the vote.
Hence, we omit all such bit strings from the analysis, as they will not contribute.
Likewise, if a variant contains no surviving bit strings, it is removed from the analysis. 
All subsequent discussions of probability are relative to this reduced collection of bit strings.

Let $f_{vb}$ be the frequency of bit string $b$ in the variant-circuit distribution indexed by $v$.
The probability that this bit string will be chosen exactly $m$ times during the process of sampling once from each of the variant distributions is given by
\begin{equation}
p_b^{(m)} = \sum_{\mathcal{S} \in \mathcal{P}_m}
\prod_{v \in \mathcal{S}} f_{vb}
\prod_{v \notin \mathcal{S}} \left(1-f_{vb}\right) , \label{eq:Gb_rel_prob}
\end{equation}
where $\mathcal{P}_m$ is the set of all $m$-element subsets of the surviving variants.
Each term in this sum can be understood as the probability of drawing $b$ in specified set $S$ of variants of size $m$, times the probability of not drawing $b$ in the remaining variants.
The sum aggregates the resulting probability over all possible sets of $m$ variants.

Given $p_b^{(m)}$, we can now compute the probability that $b$ will be chosen at least $t$ times by the sum
\begin{equation}
p_b = \sum_{m=t}^{N_v} p^{(m)}\label{eq:pb_rel_prob},
\end{equation}
where $N_v$ is the number of surviving variants.
We then normalize the set of probabilities $\{p_b : b \in \mathcal{B}\}$, with $\mathcal B$ the set of bit strings, to arrive at the final outcome distribution.

We note that, when $t \geq N_v/2$, the above algorithm calculates the exact probabilities of a majority vote above the given threshold~\cite{symm2023}. 
However, when $t < N_v/2$, the probability is approximate, since it could be the case that a given bit string occurs for more than $t$ variants, but a second bit string appears a larger number of times on the remaining variants.
This could be handled by applying an inclusion-exclusion method to Eq.~\ref{eq:Gb_rel_prob}, but here we neglect these rare events for computational simplicity.

Finally, we note that when $t < N_v/2$ (but $> 2$), it is more efficient to compute $p_b$ by summing the values of $p_b^{(m)}$ for $m < t$ and subtracting this value from 1.   Thus, when $t < N_v/2$, we replace Eq.~\ref{eq:pb_rel_prob} with
\begin{equation}
p_b = 1 - \sum_{m=0}^{t-1} G_b^{(m)}
\end{equation}
to improve algorithm efficiency.  
A Python implementation of this algorithm is provided in the supplemental data~\cite{suppdata}.

\section{M\o{}lmer–S\o{}rensen gate properties\label{sec:appendix_msgatedetails}}

As mentioned in the main text, the entangling two-qubit gates in our system are implemented with amplitude-modulated pulses designed on a per-pair basis. 
In order to achieve high fidelities in a long ion chain, we need to consider more than one normal mode of motion due to the relatively small separation between two neighboring modes, which is approximately $10$~kHz in our systems. 
All the motional modes will, to some extent, contribute to the entanglement generation unless the participant qubit happens to be at a node of the normal mode of motion; hence any ion's trajectory non-enclosure in each normal mode's phase space after applying the entangling pulse will lead to gate infidelities~\cite{choi2014optimal}.

\begin{figure}[h]
  \centering
  \includegraphics[width=\columnwidth]{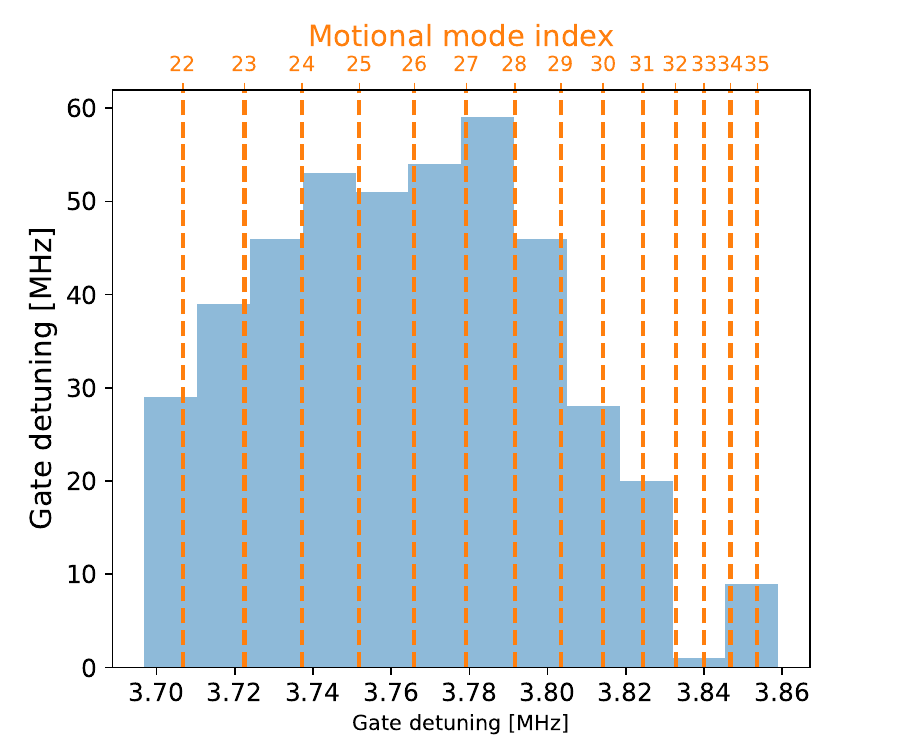}
  \caption{
  Histogram of the gate detunings of all the ${30 \choose 2} = 435$ 2-qubit gate. The vertical lines show the motional mode frequencies close to the gate detunings. Mode 35 is the center of mass mode. 
  }
  \label{fig:MSgate_nearestmodes}
\end{figure}

In order to properly close the phase space trajectories of the entangling pulses, their temporal profiles are generated via a numerical search routine like the one described in Ref.~\cite{grzesiak2020efficient}.
Besides varying the modulation profile, the optimization routine also scans over a given detuning range to search for suitable pulses that meet the given conditions such as the infidelity threshold, motional mode frequency instability, optical power limit, etc.
The resulting detunings of these pulses are distributed above modes 21, where mode 35 is the center-of-mass mode. 
The gate detuning distribution among all the pairwise MS gates with respect to the motional mode frequencies is shown in Fig.~\ref{fig:MSgate_nearestmodes}

\begin{figure}[h]
  \centering
  \includegraphics[width=\columnwidth]{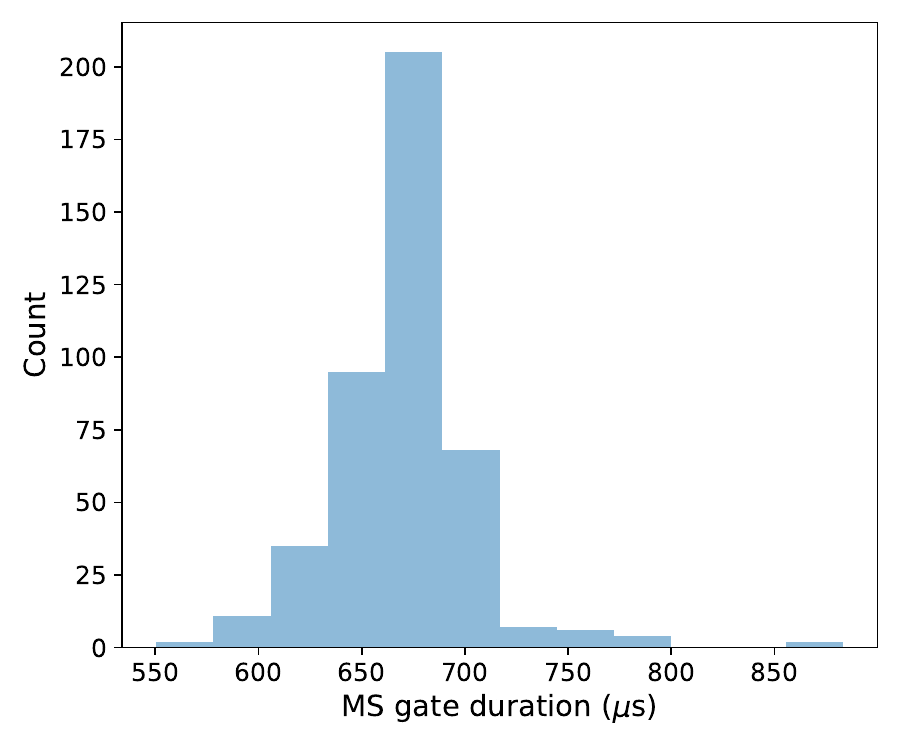}
  \caption{
  Histogram of the MS gate durations, shown for the entire set of ${30 \choose 2} = 435$ ion pairs.}
  \label{fig:MSgate_durations}
\end{figure}

The duration of each MS gate depends on the participation of normal modes, and is optimized along with other constraints.
The MS gates in Forte at the time of this writing are shown in Fig.~\ref{fig:MSgate_durations}, and range from $550$ to $883\,\mu s$ with a median value of $672\,\mu s$.  This results in ZZ gate durations between $770$ and $1103\,\mu s$.

\end{document}